\documentstyle[aps,preprint]{revtex}
\begin{document}
\preprint{\vbox{ 
\hbox{SOGANG-HEP 248/98}}}
\title{Batalin-Vilkovisky quantization of symmetric Chern-Simons theory}
\author{Won Tae Kim$^a$\footnote{electronic address: wtkim@ccs.sogang.ac.kr}, 
Chang-Yeong Lee${}^{b}$\footnote{electronic address: leecy@kunja.sejong.ac.kr},
and Dong Won Lee${}^{c}$\footnote{electronic address: dwlee@kkucc.konkuk.ac.kr}}
\address{ ${}^{a}$ Department of Physics 
and Basic Science Research Institute,\\
Sogang University, C.P.O. Box 1142, Seoul 100-611, Korea \\
$^b$ Department of Physics, Sejong University, Seoul 143-747, Korea \\ 
${}^{c}$ Department of Physics, Kon-kuk University, Seoul 143-701, Korea }
\maketitle

\begin{abstract}
We study the manifestly covariant 
three-dimensional symmetric Chern-Simons action
in terms of the Batalin-Vilkovisky quantization method. 
We find that the Lorentz covariant 
gauge fixed version of this action is reduced to the usual
Chern-Simons type action after a proper field redefinition. 
Furthermore, the renormalizability of the symmetric
Chern-Simons theory
turns out to be the same as that of the original 
Chern-Simons theory.
\end{abstract}

\newpage

Chern-Simons(CS) theory \cite{djt} 
has been studied in various arena. The
key structure which gives interesting phenomena is due to
the unusual commutator between the gauge fields, which
is essentially arisen from 
the Dirac method for the quantization of second
class constraint system \cite{dir}. 
On the other hand, the second
class constraint system can be in principle converted
into the first class constraint system by use of the 
Batalin-Fradkin-Tyutin(BFT) method \cite{bft} in the Hamiltonian
formalism. The resulting first class constraint system
is invariant with respect to the local symmetry implemented
by the first class constraints.
A few years ago, the second class constraint of
the CS theory coupled to some complex fields was converted
into first class one in the BFT Hamiltonian method \cite{ban},
and subsequently straight forward non-Abelian extension
was performed \cite{kp}. However, the Wess-Zumino like action
to convert the second class system into first class one in
the Lagrangian formulation depends on the content of 
matter couplings, and general covariance is unfortunately lost.  
 
Recently, the manifestly covariant symmetric CS action 
has been obtained \cite{ky}.
The newly obtained one has only first class constraints unlike the usual 
one which has both first class constraints and 
second class constraints. The symmetric CS theory can be obtained by simply
substituting the original gauge field in the CS action 
with the infinite sum of newly
introduced auxiliary vector fields \cite{ky}.
Note that at first sight, the appearance of the resulting 
symmetric CS action seems to be the same form
as the original CS action,
however, it is nonlocal in that the infinite series of auxiliary
fields are involved in the symmetric action. Of course, in 
the unitary gauge, the original local CS action is reproduced.

On the other hand, the Abelian CS theory 
coupled to the complex matter fields was
reconsidered in \cite{ky} as a physical application, which is
essentially first class constraint system.  
By analyzing this model without 
any gauge fixing condition, 
one can naturally obtain gauge-independent anyon operators
which are also free from path-ordering problems between field operators.
Therefore, in the symmetric formulation, the construction
of anyon operator is simply realized in the gauge-independent way 
without any ordering problems.  

In this paper, we study the symmetric CS action which has
full symmetries by use of the Batalin-Vilkovisky(BV) \cite{bv} quantization
method, and show the equivalence between the symmetric CS action and
the original CS action. 
We find that the gauge fixed version of this action turns 
out to be the same as the usual
Chern-Simons type after a proper field redefinition. 
Furthermore, the renormalizability program
turns out to be the same as that of the original 
Chern-Simons theory.

We now first recapitulate the gauge structure of the
non-Abelian CS theory. The CS action with fully first
class constraints is given as
\begin{eqnarray}
\label{scs}
{\cal S}_{\rm SCS} &=&\kappa \int d^3x  \epsilon^{\mu\nu\rho}
tr \left[ (A_{\mu} + \sum_{n=1}^{\infty} B_{\mu}^{(n)}) \partial_{\nu}
(A_{\rho} + \sum_{n=1}^{\infty} B_{\rho}^{(n)}) \right.  \nonumber \\
&-& \left. \frac{2}{3}i (A_{\mu} + \sum_{n=1}^{\infty}
B_{\mu}^{(n)}) (A_{\nu} + \sum_{n=1}^{\infty}B_{\nu}^{(n)})
(A_{\rho} + \sum_{n=1}^{\infty} B_{\rho}^{(n)}) \right],
\end{eqnarray}
where the diagonal metric $g_{\mu\nu}=diag(+,-,-)$ and
$\epsilon^{012}=+1$. The Lie algebra-valued gauge field is defined by 
$A_{\mu}=A^a_{\mu}T^a$  satisfying
$[T^a, T^b]={\it i}f^{abc}T^c$ and ${\rm tr}( T^a
T^b)=\frac{1}{2}\delta^{ab}$ where $T^a$ is a Hermitian generator. 
$A_\mu$ is an original gauge field and $B_\mu^{(n)}$ are auxiliary
vector fields introduced to make the second class constraints into
the first class constraints \cite{ky}.
The Lagrangian is invariant under the following gauge
transformations up to a total divergence,
\begin{eqnarray}
\label{transformation}
\delta A_{\mu} &=& D_{\mu} \epsilon^{(0)}+ \left[\sum_{n=1}^{\infty}
B_{\mu}^{(n)},~~\epsilon^{(0)}\right] + \epsilon_{\mu}^{(1)}, \nonumber \\
\delta B_{\mu}^{(n)} &=&
-\epsilon_{\mu}^{(n)}+\epsilon_{\mu}^{(n+1)}  ~~~   (n = 1,2,\cdots) ,
\end{eqnarray}
where $\epsilon^{(0)}$ and $\epsilon^{(n)}$ are 
independent parameters of the local symmetries and
$D_{\mu}= \partial_{\mu} + [ A_{\mu}, ~~ ]$. 
Considering the commutator of
two gauge transformations of the above type (\ref{transformation}), 
we see that
\begin{equation}
[ \delta_1 , \delta_2] A_{\mu}^a = D_{\mu}(f^a_{bc}
\epsilon_1^{(0)b} \epsilon_2^{(0)c})
+ f^a_{bc} B_{\mu}^{(n) b} f^c_{de} \epsilon_1^{(0)d}
\epsilon_2^{(0)e},
\end{equation}
\begin{equation}
[\delta_1 , \delta_2] B_{\mu}^{(n)a} = 0 ,
\end{equation}
and thus
$\epsilon^{(0)a}_{12}=f^a_{b c}\epsilon_1^{(0)b}\epsilon_2^{(0)c}$.
These relations tell us that 
only the gauge parameter $\epsilon^{(0)}$ has the original group
structure, and the symmetry algebra is closed and irreducible.

To quantize the symmetric CS action (\ref{scs}), 
we impose the restriction to the space of all
histories in order to get the constrained surface $\Sigma$.
In the BV formalism \cite{bv}, an
 antifield $\Phi^{\ast}$ for each field $\Phi$ is introduced to implement
 this procedure.
In our case, $\Phi$ includes the gauge fields $A_{\mu}^a$, $B_{\mu}^{(n)a}$
as well as the ghost fields $C^a$ and $C^{(n)}_a$ which are
corresponding to the gauge parameters
$\epsilon^{(0)}_a$ and $\epsilon^{(n)}_a$, respectively,
\begin{eqnarray}
\Phi^A &=& \left( A_{\mu}^a, B_{\mu}^{(n) a}, C^a, C_{\mu}^{(n) a} \right)
,                                           \nonumber \\
\Phi^{\ast}_A &=& \left( A_{\mu a}^{\ast}, B_{\mu a}^{(n)\ast},
C^{\ast}_a, C_{\mu a}^{(n)\ast} \right).
\end{eqnarray}
The ghost number and statistics of $\Phi_A^{\ast}$ are
assigned as
\begin{eqnarray}
gh[\Phi^{\ast}_A] &=& -gh[\Phi^A]-1 ,           \nonumber \\
\epsilon(\Phi_A^{\ast}) &=& \epsilon(\Phi^A) +1 (\mbox{mod}~2) ,
\end{eqnarray}
such that the statistics of $\Phi_A^{\ast}$ is opposite to that of
$\Phi^A$. Then the anti-bracket is defined by
\begin{equation}
(X, Y) \equiv \frac{\partial_r X}{\partial \Phi^A}\frac{\partial_l
Y}{\partial \Phi_A^{\ast}} - \frac{\partial_r X}{\partial
\Phi_A^{\ast}} \frac{\partial_l Y}{\partial \Phi^A} .
\end{equation}
In the BV formalism, the action $S[\Phi ,\Phi^{\ast}]$ 
should be a
functional of fields and antifields satisfying the master equation,
\begin{equation}
(S,S)=2\frac{\partial_r S}{\partial \Phi^A} \frac{\partial_l
S}{\partial \Phi^{\ast}_A} =0.
\label{masteq}
\end{equation}
The solution $S$ of the master equation can be expanded
in a power series in antifields. 
In our case, it has non-vanishing structure
constants only up to the first order, and 
the minimal solution  for 
the master equation can be written as
\begin{eqnarray}
S_{\rm Min}= && \int d^3x \frac{\kappa}{2} \epsilon^{\mu\nu\rho}
\left[(A_{\mu} + \sum_{n=1}^{\infty} B_{\mu}^{(n)})^a \partial_{\nu}
(A_{\rho} + \sum_{n=1}^{\infty} B_{\rho}^{(n)})^a \right.  \nonumber \\
&+&\left. \frac{1}{3} f^a_{b c}(A_{\mu} + \sum_{n=1}^{\infty}
B_{\mu}^{(n)})^a (A_{\nu} + \sum_{n=1}^{\infty}B_{\nu}^{(n)})^b
(A_{\rho} + \sum_{n=1}^{\infty} B_{\rho}^{(n)})^c \right] \nonumber\\
&+& \int d^3x \left[A_{\mu a}^{\ast} (D^{\mu} C^a +f^a_{bc}
\sum_{n=1}^{\infty} B_{\mu}^{(n) b} C^c +C_a^{(1) \mu})\right. \nonumber
\\  &+&  \sum_{n=1}^{\infty} B_{\mu a}^{(n) \ast}(-C^{(n) \mu}_a
+C^{(n+1) \mu}_a) + \left. \frac{1}{2}f^a_{bc} C^{\ast}_a
C^b C^c \right] ~~~.
\end{eqnarray}
At this stage, the BRST variation of a functional $X$ is given by
the anti-bracket with the minimal action
\begin{equation}
\delta_B X \equiv (X, S_{\rm Min}),
\end{equation}
and thus we obtain the BRST transformations for fields and antifields 
as follows. 
\begin{eqnarray}
\label{brst}
\delta_B A^{\mu}_a &=& D^{\mu} C^a + f^a_{bc} \sum_{n=1}^{\infty}
B_{\mu}^{(n) b} C^c +C_a^{(1) \mu},
                                                  \nonumber \\
\delta_B A^{\ast a}_{\mu} &=& -\frac{\partial_l S_0}{\partial
A^{\mu}_a} -f^a_{bc} A^{\ast c}_{\mu} C^b ,       \nonumber \\
\delta_B B^{(n) \mu}_a &=& -C^{(n) a}_{\mu} + C^{(n+1) a}_{\mu}
,                                              \nonumber \\
\delta_B B^{(n) \ast a}_{\mu} &=& -\frac{\partial_l S_0}{\partial
B^{(n)\mu}_a} -f^a_{bc} A^{\ast c}_{\mu} C^b
 ,                                                        \\
\delta_B C^a &=& \frac{1}{2}f^a_{bc} C^b C^c ,       \nonumber  \\
\delta_B C^{\ast}_a &=& -D^{\mu} A^{\ast a}_{\mu} + f^a_{bc}
B^{(n)\mu b} A^{\ast c}_{\mu} -f^a_{bc} C^{\ast}_b C_c
,                                               \nonumber \\
\delta_B C^{(n) \mu}_a &=& 0 ,                      \nonumber \\
\delta_B C^{(n) \ast}_{\mu a} &=& B^{(n-1) \ast}_{\mu a} -B^{(n)
\ast}_{\mu a} ~,                  \nonumber
\end{eqnarray}
where $ B^{(0) \ast}_{\mu a} = A^{\ast}_{\mu a} $.
One can check that with the above transformation (\ref{brst}) 
the minimal action 
satisfies the master equation:
\begin{equation}
\label{master}
(S_{\rm Min} , S_{\rm Min}) =0.
\end{equation}

We are now in a position to fix a gauge, 
and to do that we add an auxiliary action which is a trivial-pair 
type solution of the master equation,
\begin{equation}
S_{\rm Aux} = \int d^3x (\bar{\pi_a} \bar{C_a}^{\ast} + \bar{\pi_a}^{(1)\mu}
\bar{C_a}^{(1)\ast}_{\mu} + \cdots \cdots +\bar{\pi_a}^{(n)\mu}
\bar{C_a}^{(n) \ast}_{\mu} +\cdots) .
\end{equation}
Obviously, the combined action, which we will call non-minimal, 
\begin{equation}
S_{\rm NM} = S_{\rm Min} + S_{\rm Aux},
\end{equation}
satisfies the master equation (\ref{masteq}), and
contains the non-minimal
set of fields ($\bar{\pi} , \bar{\pi}^{(n)} , \bar{C} ,\bar{C}^{(n)}$).

For the elimination of antifields, 
we choose the so-called 
gauge-fixing fermion
$\Psi$ as
\begin{eqnarray}
\Psi &=& \int d^3 x \left[ \bar{C}_a \partial^{\mu} (A_{\mu} +
\sum_{n=1}^{\infty} B^{(n)}_{\mu})^a +\sum_{n=1}^{\infty}
\bar{C}^{(n)}_{\mu a} B^{(n)\mu a} \right]               \nonumber \\
&-& \frac{1}{2}\int d^3x \left[ \frac{\bar{C}_a \bar{\pi}_a}{\xi_0} +
\frac{\bar{C}^{(1)}_{\mu a} \bar{\pi}^{(1)\mu}_a}{\xi_1} + \cdots
\cdots + \frac{\bar{C}^{(n)}_{\mu a} \bar{\pi}^{(n)\mu}_a}{\xi_n}+
\cdots \right],
\label{gffermion}
\end{eqnarray}
which is {\it admissible} \cite{ht}, so that the theory becomes non-degenerate.
Note that the above choice of the gauge fixing fermion $\Psi$ corresponds
to two types of gauge fixing conditions for the fields $A_{\mu}$ and 
$ B^{(n)}_{\mu}$:
\begin{eqnarray}
U^{(0)} &=& \partial^{\mu} (A_{\mu} +
\sum_{n=1}^{\infty} B^{(n)}_{\mu}) , \nonumber \\
U^{(n)} &=& B^{(n)}_\mu, ~~~~~ n \geq 1 .
\label{gaugefix}
\end{eqnarray}
The antifields $\Phi^{\ast}$ are
eliminated by the relation $\Phi^{\ast} = \frac{\partial \Psi}
{\partial \Phi}$, and our choice of $\Psi$ yields the following relations.
\begin{eqnarray}
A^{\ast a}_{\mu} &=& - \partial_{\mu} \bar{C}^a ,\nonumber \\
B^{(1)\ast a}_{\mu} &=& -\partial_{\mu} \bar{C}^a +\bar{C}^{(1) a
}_{\mu}  ,                                    \nonumber \\
B^{(n)\ast a}_{\mu} &=& -\partial_{\mu} \bar{C}^a
+\bar{C}^{(n)a}_{\mu} ,                      \nonumber \\
\bar{C}^{\ast}_a &=& \partial^{\mu}(A_{\mu} + \sum_{n=1}^{\infty}
B^{(n)}_{\mu})^a -\frac{1}{2}\frac{\bar{\pi}}{\xi_0} ,  \nonumber \\
\bar{C}^{(n) \ast a}_{\mu} &=& B^{(n)}_{\mu}
-\frac{1}{2}\frac{\bar{\pi}^{(n)}_{\mu}}{\xi_n} , \nonumber \\
C^{\ast} &=& 0 .
\end{eqnarray}
Plugging these into the non-minimal action $S_{\rm NM}$ and
after performing Gaussian integrations over the field $\bar{\pi}$ and
$\bar{\pi}^{(n)}$, we obtain the following gauge-fixed action,
\begin{eqnarray}
S_{\Psi} =  && \int d^3x \frac{\kappa}{2} \epsilon^{\mu\nu\rho}
\left[(A_{\mu} + \sum_{n=1}^{\infty} B_{\mu}^{(n)})^a \partial_{\nu}
(A_{\rho} + \sum_{n=1}^{\infty} B_{\rho}^{(n)})^a \right.  \nonumber \\
&+&\left. \frac{1}{3} f^a_{b c}(A_{\mu} + \sum_{n=1}^{\infty}
B_{\mu}^{(n)})^a (A_{\nu} + \sum_{n=1}^{\infty}B_{\nu}^{(n)})^b
(A_{\rho} + \sum_{n=1}^{\infty} B_{\rho}^{(n)})^c \right] \nonumber\\
 &+& \int d^3x \left[ \bar{C}_a \partial_{\mu}(D^{\mu} C^a
+f^a_{bc} \sum_{n=1}^{\infty} B_{\mu}^{(n) b} C^c) \right.
                                                   \nonumber \\
&+& \sum_{n=1}^{\infty} \bar{C_{\mu}}^{(n)}_a (-C^{(n)\mu} +
C^{(n+1)\mu})_a + \frac{\xi_0}{2}(\partial^{\mu} (A_{\mu} +
\sum_{n=1}^{\infty} B^{(n)}_{\mu}))^2               \nonumber \\
 &+& \left. \sum_{n=1}^{\infty} \frac{\xi_{(n)}}{2}(B_{\mu}^{(n)})^2 \right] .
\label{gfaction}
\end{eqnarray}
One thing we have to note is that the (anti)ghost fields $\bar{C_{\mu}}^{(n)}$ 
and $C^{(n)}_{\mu}$ do not have kinetic terms and these fields simply
provide delta function relations among $C^{(n)}_{\mu}$'s, 
$ \delta ( C^{(n)}_{\mu} - C^{(n+1)}_{\mu})$, 
if we integrate over $\bar{C_{\mu}}^{(n)}$. 

Now, the BRST transformation of a functional $X$ 
after gauge fixing is given by the anti-bracket with 
the non-minimal action $S_{NM}$ restricted on $\Sigma_{\Psi}$
\begin{equation}
\delta_{B_{\Psi}} X \equiv (X,S_{NM})\mid_{\Sigma_{\Psi}} ,
\end{equation}
where $\Sigma_{\Psi}$
denotes the constraint surface determined by the condition
\begin{equation}
Y(\Phi,\Phi^{\ast})\mid_{\Sigma_{\Psi}} \equiv Y (\Phi, \frac{\partial \Psi}{\partial
\Phi}) .
\end{equation}
Thus, the final BRST transformation after gauge fixing is given by 
\begin{eqnarray}
\label{newbrst}
\delta_{B_{\Psi}} A^{\mu}_a &=& D^{\mu} C^a + f^a_{bc} \sum_{n=1}^{\infty}
B_{\mu}^{(n) b} C^c +C_a^{(1) \mu} ,      \nonumber \\
\delta_{B_{\Psi}} B^{(n) \mu}_a &=& -C^{(n) a}_{\mu} + C^{(n+1) a}_{\mu}
,      \nonumber \\
\delta_{B_{\Psi}} C^a &=& \frac{1}{2}f^a_{bc} C^b C^c ,   \nonumber  \\
\delta_{B_{\Psi}} C^{(n) \mu}_a &=& 0   ,       \\
\delta_{B_{\Psi}} \bar{C}^a &=& \xi_0 \partial^{\mu} (A_{\mu} +
\sum_{n=1}^{\infty} B^{(n)}_{\mu})^a  ,     \nonumber \\
\delta_{B_{\Psi}} \bar{C}_{\mu}^{(n) a} &=& \xi_{(n)} B_{\mu}^{(n) a} .
\nonumber
\end{eqnarray}
One can again check that the gauge fixed action $S_{\Psi}$
is invariant under the above BRST 
transformation (\ref{newbrst}).
\\

We now turn to the renormalizability of the theory.
To find the propagators,
we first express the
quadratic part of the gauged fixed action (\ref{gfaction}) for the fields
$A_{\mu}$, $B_{\mu}^{(n)}$,
\begin{eqnarray}
 \begin{array}{cccccc}
 \ & A_{\rho} & B_{\rho}^{(1)} & B_{\rho}^{(2)} & B_{\rho}^{(3)} & \cdots  \cr
 A_{\mu}   &     C       &        C       &    C     &    C    & \cdots  \cr
 B_{\mu}^{(1)} &    C    &      C+D_1     &    C     &    C    & \cdots  \cr
 B_{\mu}^{(2)} &    C    &        C       &  C+D_2   &    C    & \cdots  \cr
 B_{\mu}^{(3)} &    C    &        C       &    C     &  C+D_3  & \cdots  \cr
        .  &    .        &        .       &    .     &    .    &    .     \cr
    \end{array}
\end{eqnarray}
where $C  =  \kappa \epsilon^{\mu \nu \rho} P_{\nu}
-\xi_0P^{\mu}P^{\rho},~ D_n  =  -\xi_n g^{\mu \rho}$.
After diagonalization, this can be written as
\begin{eqnarray}
 \begin{array}{cccccc}
 \ & \bar{A}_{\rho}&B_{\rho}^{(1)}&B_{\rho}^{(2)}&B_{\rho}^{(3)}&\cdots  \cr
\bar{A}_{\mu}  &    C    &        0       &    0     &    0    & \cdots  \cr
 B_{\mu}^{(1)} &    0    &        D_1     &    0     &    0    & \cdots  \cr
 B_{\mu}^{(2)} &    0    &        0       &    D_2   &    0    & \cdots  \cr
 B_{\mu}^{(3)} &    0    &        0       &    0     &    D_3  & \cdots  \cr
        .  &    .        &        .       &    .     &    .    &    .     \cr
    \end{array}
\end{eqnarray}
where $\bar{A}_{\mu}$ is defined as 
\begin{equation}
\bar{A}_{\mu} \equiv A_{\mu} + \sum_{n=1}^{\infty} B_{\mu}^{(n)}.
\end{equation}

If we set the dimensionless parameter $\kappa=1$, then we obtain the 
following propagator 
for the field $\bar{A}_{\mu}$.
\begin{equation}
\triangle_{\mu\nu}^{ab} =\frac{1}{p^2} \left[ \epsilon_{\mu\nu\rho}
p^{\rho} -\frac{p_{\mu}p_{\nu}}{\xi_0 p^2} \right] \delta^{ab}.
\end{equation}
For the ghost field
$C^a$, the propagator is given by
\begin{equation}
\Lambda_{ab} = \frac{1}{p^2} \delta_{ab}.
\end{equation}
The propagators for $B_{\mu}^{(n)}$ are trivial and decouples from the
theory.
Remember that integrating out $\bar{C_{\mu}}^{(n)}$ in the gauge fixed 
action (\ref{gfaction}) gives the
delta function relations among $ C_{\mu}^{(n)}$'s. Thus  
 after the field redefinition these relations tell us that the BRST 
 variations of $B_{\mu}^{(n)}$ are vanishing. Also, the BRST 
 variation of $\bar{A_{\mu}}$ becomes the usual one,
 \[ \delta \bar{A_{\mu}}^a = \partial_{\mu} C^a + f^a_{bc}\bar{A_{\mu}}^b C^c, \] 
 whereas the original variation of $A_{\mu}$
(\ref{newbrst}) before the field redefinition contains an 
extra ghost field $C^{(1)}_{\mu}$.
Thus the propagating fields are only $\bar{A_{\mu}}, ~~ C^a$, and $\bar{C}^a , ~$
just like the original CS theory. 
Furthermore, the contributing propagators are the same as in the usual CS theory 
which had been investigated and shown to be one-loop renormalizable \cite{brt,alr}.
Therefore, we can conclude that our generalized first-class action 
has the same renormalizability
property as the usual CS action and hence one-loop
renormalizable.

This result is somewhat expected, because 
our starting symmetric CS action (\ref{scs}) is designed to
maintain the local physical properties in the enlarged configuration
space. After all, it should be possible that the enlarged 
first class system be gauged away by a certain   
gauge condition, and our gauge condition (\ref{gaugefix}) leads to the
same physics as that of the usual CS theory.
 
As a comment, one might be wonder why the form of 
the symmetric action (\ref{scs}) which
is fully first class constraint system is the same as
that of the original CS action when one defines
${\bar A}_\mu =A_{\mu} + \sum_{n=1}^{\infty} B_{\mu}^{(n)}$. 
That is, the second class
constraint algebra seems to appear again in the 
symmetric action case, if one regards $\bar{A_\mu}$ as a fundamental
field. However, this is not the case since
${\bar A}_\mu$ is not a fundamental local field but 
composed of infinite number of vector fields. Therefore,
we should note that the starting action (\ref{scs})
is in some sense a nonlocal action. 
Unfortunately, we do not know at this stage 
how to convert the second class
constraint system of CS action into the first class constraint
system by introducing only finite number
of auxiliary fields.

In summary, we quantized 
the symmetric CS action in 
the BV formalism. In the symmetric CS theory, the
auxiliary vector fields $B^{(n)}_\mu$ can be naturally eliminated 
after diagonalizing the quadratic part of the action.
The propagator of the 
nonlocal vector field ${\bar A_\mu}$ which involves the
infinite number of auxiliary vector fields is remarkably
written as the same form as that of the pure CS theory. 
As a result, it is equivalent to the original
CS theory under the proper gauge condition, and renormalizable
similarly to the original CS action.

\acknowledgements
{W.T.K. was supported in part by
the Basic Science Research Institute Program, the Ministry of
Education, Project No. BSRI-98-2414, and 
C.Y.L. and D.W.L. were supported
in part by the Ministry of
Education, BSRI-98-2442, and KOSEF Grant 971-0201-007-2.}

\end{document}